\documentclass[letterpaper, 10 pt, conference]{ieeeconf}  

\IEEEoverridecommandlockouts                              

\usepackage{xcolor}
\usepackage{caption}
\usepackage{amssymb}
\usepackage{graphicx}
\usepackage{caption}
\usepackage{algorithm}
\usepackage{mathtools}
\usepackage{algorithmicx}
\usepackage{algpseudocode} 
\usepackage{amsmath}
\usepackage{booktabs}
\usepackage{subcaption}
\usepackage{comment}
\usepackage{wrapfig}
\usepackage{float}
\usepackage{bm}
\usepackage{setspace}

\newtheorem{rem}{Remark}
\usepackage{url}

\title{\LARGE \bf
Signal Temporal Logic Evaluation and Synthesis Using  \\ Deep Reachability Analysis and Layered Control Architecture
}

\author{Joonwon Choi$^{1}$, Kartik Anand Pant$^{1}$, Youngim Nam$^{2}$, Henry Hellmann$^{1}$, Karthik Nune$^{3}$, and Inseok Hwang$^{1}$%
\thanks{$^{1}$The authors are with School of Aeronautics and Astronautics, Purdue University, West Lafayette, IN 47907, USA
        {\tt\small choi774, kpant, hhellman, ihwang@purdue.edu}}%
\thanks{$^{2}$The author is with Department of Mechanical Engineering, Ulsan National Institute of Science and Technology, Ulsan, 44919, Republic of Korea
        {\tt\small nyi0944@unist.ac.kr}}%
\thanks{$^{3}$The author is with Interdisciplinary Engineering, Purdue University, West Lafayette, IN 47907, USA
        {\tt\small knune@purdue.edu}}%
}

\begin{document}

\maketitle
\thispagestyle{empty}
\pagestyle{empty}


\begin{abstract}

We propose a signal temporal logic (STL)-based framework that checks the feasibility of a mission described in STL and synthesizes control to safely execute it. The proposed framework evaluates the safety and reliability of operation through two phases. First, the proposed framework assesses the feasibility of STL by computing a backward reachable tube (BRT), which captures all states that can satisfy the given STL. The proposed framework accommodates the multiple reach-avoid (MRA) problem to address more general STL specifications and leverages a deep neural network to alleviate the computation burden for reachability analysis, significantly reducing the inference time compared to a baseline method. 
We further propose a layered planning and control architecture that combines mixed-integer linear programming (MILP) for global planning with model predictive control (MPC) as a local controller for the given STL. Consequently, the proposed framework can robustly handle unexpected behavior of obstacles that are not described in the environment information or STL, thereby providing reliable mission performance.
Our numerical simulations demonstrate that the proposed framework can successfully compute BRT for a given STL and perform the mission.

\end{abstract}

\section{INTRODUCTION}

Signal temporal logic (STL) is a specification language that can express a given specification of a target dynamical system using a combination of logic and temporal operators \cite{chen2018signal}.
STL can explicitly consider real-valued variables and dense-time requirements~\cite{maler2004monitoring} and thus can precisely represent a specification in a formal manner. Thanks to this benefit, STL has been widely used by several machine learning algorithms to represent spatio-temporal constraints or to enhance explainability~\cite{leung2019backpropagation, decastro2020interpretable, leung2023backpropagation, puranic2021learning, meng2023signal}, along with other temporal logic approaches. For instance, a linear temporal logic (LTL)-based method was proposed to monitor the safety of LLM-based planner~\cite{wang2024ensuring}, whereas minimal relaxation for STL-based planning was proposed in~\cite{buyukkocak2025resilient}. 

While STL feasibility can be verified by MILP optimization (e.g., Gurobi returning an irreducible inconsistent subset (IIS), ~\cite{ghosh2016diagnosis}), it remains point-based and only considers a given (set of) initial state of the system. In contrast, reachability analysis considers all the reachable states of a target system, providing set-based analysis results ~\cite{jiang2024guaranteed,arfvidsson2024ensuring}. Particularly, \cite{chen2018signal} demonstrated that STL satisfaction can be examined using the backward reachable tube (BRT). However, general reachability analysis typically considers a single target-constraint set and thus, one might not be able to rigorously analyze STL involving multiple intermediate missions. Furthermore, computing BRT involves solving the Hamilton-Jacobi (HJ) PDE, suffering from the curse of dimensionality. To tackle this problem, deep learning-based methods have been actively investigated to alleviate the high computation burden~\cite{bansal2021deepreach}. 

Meanwhile, even if the STL itself is verified to be feasible, unexpected environmental changes or incorrect information may cause problems. For instance, there may be an unobserved obstacle, or the target state may differ from the known information. To address this issue, several algorithms have been proposed to fix STL specifications or to synthesize safe control inputs~\cite {mallozzi2022contract, buyukkocak2022temporal}. 

Motivated by the aforementioned problems, we propose an end-to-end STL evaluation and control synthesis framework that efficiently confirms the feasibility and robustly executes a given STL mission. The proposed framework can be divided into two parts: \textit{DeepSTLReach} for STL evaluation and \textit{Layered-architecture} for planning and control. For specific types of STL specifications, \textit{DeepSTLReach} provides a \textit{set of initial states} that can satisfy a given STL specification, considering the capability of a target dynamical system. Thus, one can obtain more comprehensive analysis results than evaluating a specific initial state or trajectory. 
Moreover, \textit{DeepSTLReach} addresses the multiple reach-avoid (MRA) problem~\cite{chen2025control, chen2025control_arxiv} to resolve the conflict due to multiple unordered target-constraint sets of STL specifications. Interpreting STL from a reachability perspective may yield multiple target sets without a specific order for visiting~\cite {chen2018signal}, thereby often making it difficult to formulate as a standard reachability analysis. On the other hand, MRA considers the reachability for visiting a sequence of multiple target-constraint sets. Accordingly, one can examine the feasibility of a broader range of STL specifications.
Furthermore, to reduce computational time for BRT generation, we employ a deep learning-based approach. Motivated by \cite{bansal2021deepreach}, we train a neural network to obtain BRT through supervised learning. Consequently, the proposed framework can compute BRT significantly faster, making it suitable for online monitoring of a dynamical system.

Subsequently, once we evaluate that the given STL is feasible, we formulate a trajectory-planning and tracking problem using a layered control architecture \cite{matni2024towards} combining MILP and model predictive control (MPC), respectively. STL specifications are transformed into integer constraints via robustness, which serves as a measure of how well the constraints are satisfied. Using the MILP-based planning as the reference trajectory, we employ an MPC-based tracking controller that tracks the reference trajectory while locally enforcing safety constraints (e.g., in a warehouse, the position of some of the obstacles may change during the operations, making the previous STL infeasible. The MPC-based controller is leveraged to handle these runtime safety violations).

In summary, our main contributions in this paper are:
1) We propose \textit{DeepSTLReach} to evaluate the feasibility of STL using reachability analysis. \textit{DeepSTLReach} can address multiple unordered target-constraint sets that might be inherent within the given STL specification by leveraging MRA. Thus, one can evaluate a broader range of STL specifications while considering the physical capability of a dynamical system; 2) We leverage deep reachability analysis to reduce the computational complexity to compute BRT. As a result, \textit{DeepSTLReach} significantly reduces the inference time compared to a baseline algorithm; 3) We propose a layered planning and control architecture combining MILP for global planning, enforcing spatio-temporal constraints via STL specifications, and MPC-based local planning and control to perform a mission considering runtime safety; and 4) We rigorously test the performance of the proposed framework through a set of illustrative numerical simulations.

The rest of the paper is organized as follows. In Section \ref{prelm}, the preliminaries on STL and reachability analysis are presented. Section \ref{proposed_framework} provides a detailed description of our proposed framework. In Section \ref{sec:experiment}, the results of the numerical simulations and experiments are presented. Lastly, the conclusion is given in Section \ref{sec:conclusion}.

\section{Preliminaries} \label{prelm}
Let the dynamics of a target dynamical system be: 
\begin{equation} \label{dynamics}
    \dot{x}=f(x,t,u,d)
\end{equation}
where $x\in\mathbb{R}^n$ is the state, $u\in\mathbb{R}^m$ is the control input, $d\in\mathbb{R}^e$ is the disturbance, $t$ is time, and $f(\cdot)$ is the dynamic model of the system.

\subsection{Signal temporal logic (STL)}
The syntax of STL can be represented using $\text{T}\ |\ P \ |\ \neg\phi \ |\ \phi_1\wedge\phi_2\ |\ \phi_1 \mathcal{U}_{[a,b]}\phi_2$, where $\phi_{(\cdot)}$ is STL formula, $\text{T}$ is logical true, $P$ is a predicate, $\neg$ is Boolean negation, $\wedge$ is conjunction, and $\mathcal{U}_{[a,b]}$ is the $\textit{until}$ operator with $a\leq b\in\mathbb{R}$. The other operators also can be represented using the aforementioned syntax such as $\textit{disjunction}\ (\vee, \neg(\neg\phi_1 \wedge \neg\phi_2))$, $\textit{eventually}\ (\Diamond, \text{T}\mathcal{U}_{[a,b]}\phi)$, and $\textit{always}\ (\square, \neg \Diamond_{[a,b]}\neg\phi)$~\cite{leahy2023rewrite}. 
The satisfaction of such an STL can be represented using $\textit{robustness},\ \rho(x,t,\phi)$, a quantitative semantics of STL~\cite{donze2010robust}, which is defined as:
\begin{align}
&\rho(x,t,\pi(x(t))\geq c) := \pi(x(t))-c  \label{stl_pred}\\
&\rho(x,t,\neg \phi) := -\rho(x,t,\phi) \label{stl_neg} \\
&\rho(x,t,\phi_1 \wedge \phi_2) := min(\rho(x,t,\phi_1),\rho(x,t,\phi_2)) \label{stl_and} \\
&\rho(x,t,\phi_1 \mathcal{U}_{[a,b)}\phi_2) := \nonumber \\ & \ \ \ \ \ \ \max_{t'\in[t+a,t+b]} \min (\rho(x,t',\phi_2),\min_{t''\in[t,t']} \rho(x,t'',\phi_1)) \label{stl_until}
\end{align}
with respect to the state $x$ at time $t$. If $\rho(x,t,\phi)>0$, this means STL $\phi$ is satisfied at the state $x$ and time $t$. Unless stated otherwise, we write $\rho(x,0,\phi)$ as $\rho_{\phi}(x)$ for brevity throughout the paper.

\subsection{Backward reachable tube (BRT)}
To evaluate the feasibility of an STL, we utilize BRT that captures all states that the target system can reach to a target within an arbitrary time horizon. 
The maximal BRT $\\ \mathcal{R}^M(t,T;\mathcal{T}(\cdot),\mathcal{C}(\cdot))$, of \eqref{dynamics} is defined as~\cite{chen2018signal}:
{\small \begin{multline} \label{brt}
\mathcal{R}^M(t,T;\mathcal{T}(\cdot),\mathcal{C}(\cdot))=\{x:\exists u(\cdot) \in \mathbb{U}, \forall d(\cdot) \in \mathbb{D}, \exists s \in[t,T], \\ \xi_{x,t}^{u(\cdot),d(\cdot)}(s)\in\mathcal{T}(s), \forall \tau\in[t,s], \xi_{x,t}^{u(\cdot),d(\cdot)}(\tau)\in\mathcal{C}(\tau) \}
\end{multline}}
where $ \xi_{x,t}^{u(\cdot),d(\cdot)}$ is the trajectory that starts from state $x$ and time $t$ following the control input $u(\cdot)$ with disturbance $d(\cdot)$. $\mathbb{U}$ and $\mathbb{D}$ denote the spaces of all admissible control inputs and disturbances, respectively; $\mathcal{T}$ is the target set; and $\mathcal{C}$ is the constraint set. 
$\mathcal{R}^M$ is a sublevel set of the Bellman value function, i.e., $\mathcal{R}^M(t,T;\mathcal{T}(\cdot),\mathcal{C}(\cdot)) = \{x:h_{\mathcal{R}^M}(t,x) <0\}$ where $h_{\mathcal{R}^M}(t,x)$ is defined as follows for given $h_\mathcal{T}(t,x)$ and $  h_\mathcal{C}(x)$ of the target and constraint, respectively~\cite{chen2018signal}:
\begin{multline} \label{Bellman}
    h_{\mathcal{R}^M} (t,x)= \inf_{u(\cdot)\in\mathbb{U}} \sup_{d(\cdot)\in\mathbb{D}} \min_{s\in[t,T]} \max \{h_\mathcal{T} (t, \xi_{x,t}^{u(\cdot),d(\cdot)}(s)), \\\max_{\tau\in[t,s]} h_\mathcal{C}(\xi_{x,t}^{u(\cdot),d(\cdot)}(\tau)) \}
\end{multline}
$h_{\mathcal{R}^M}(t,x)$ can be computed by solving the Hamilton–Jacobi Variational Inequality (HJ-VI). Similarly, the minimal BRT can be defined as 
\begin{multline} \label{minimal_brt}
\mathcal{R}^m(t,T;\mathcal{T}(\cdot))=\{x:\forall u(\cdot) \in \mathbb{U}, \exists d(\cdot) \in \mathbb{D}, \exists s \in[t,T], \\ \xi_{x,t}^{u(\cdot),d(\cdot)}(s)\in\mathcal{T}(s)\}.
\end{multline}
Similar to $\rho_{\phi}$, we write $h_{(\cdot)}(0,x)$ as $h_{(\cdot)}(x)$ for brevity throughout the paper.

In~\cite{chen2018signal}, BRT is investigated to compute the set of states that can satisfy given STL specification.
Let $\mathcal{S}_{\phi}$ be a set of states that satisfy the STL $\phi$. In other words, $(\xi_{x,t}^{u(\cdot),d(\cdot)}(\cdot),s) \models \phi \Leftrightarrow \xi_{x,t}^{u(\cdot),d(\cdot)}(s) \in \mathcal{S}_{\phi}$.
From the definition of $\rho$, it is clear that $\mathcal{S}_{\phi}:=\{x :\rho_{\phi}(x)>0 \}$. 
Then, the robustness ($\rho$) and Bellman value function of $\mathcal{S}_{\phi}$ ($h_{\mathcal{S}_{\phi}}$) have the following correspondence:
\begin{equation} \label{corresp}
    \rho_{\phi}(x) > 0 \Leftrightarrow h_{\mathcal{S}_{\phi}}(x)<0
\end{equation}
In other words, a feasible set of states that satisfy the given STL $\phi$ ($\mathcal{S}_{\phi}$) can be computed using a reachability formulation.
Intuitively, one can compute the set of initial states in which a target system can accomplish a mission described by STL ($\phi$) by computing \eqref{brt}. 

Moreover, based on \eqref{stl_pred}-\eqref{stl_until} \textit{and} ($\wedge$) and \textit{or} ($\vee$) satisfy the followings relationship: $ h_{\mathcal{S}_{\phi_1\wedge\phi_2}}(x)=max(h_{\mathcal{S}_{\phi_1}}(x),h_{\mathcal{S}_{\phi_2}}(x))$ and $h_{\mathcal{S}_{\phi_1\vee\phi_2}}(x)=min(h_{\mathcal{S}_{\phi_1}}(x),h_{\mathcal{S}_{\phi_2}}(x))$~\cite{chen2018signal}.
Such a correspondence between STL and BRT can further express various operators, including \textit{until}, \textit{eventually}, and \textit{always}. 
For \textit{until} operator, $\phi_1 \mathcal{U}_{[t_1,t_2]}\phi_2$, if we define a target set and constraint set as
\begin{align} \label{until}
 &    \mathcal{T}_{\phi_2}= 
\begin{cases} 
    \{x:(x(\cdot),s) \models \phi_2 \}, & \text{if } s\in[t+t_1, t+t_2] \\
    \emptyset & \text{otherwise, }
\end{cases} \\
& C_{\phi_1} = \{x:(x(\cdot),s) \models \phi_1 \} \nonumber,
\end{align}
the corresponding BRT can be defined as $\mathcal{R}^M(t,T,\mathcal{T}_{\phi_2}(\cdot),  C_{\phi_1}(\cdot))$ from \eqref{Bellman}, i.e., one can compute $h_{\mathcal{R}^M}=h_{\mathcal{S}_{\phi_1 \mathcal{U}_{[t_1,t_2]}\phi_2}}$. 
\textit{Eventually} can be computed by setting $\phi_1 = True$, i.e., $\Diamond_{[t1,t2]}\phi= True \  \mathcal{U}_{[t_1,t_2]}\phi$.

\textit{Always} requires the minimal BRT computation \eqref{minimal_brt}. For $\square_{[t1,t2]}\phi$, one can set 
\begin{align}
    & \mathcal{T}_{\phi}= \label{always_target}
\begin{cases} 
    \{x:(x(\cdot),s)  \not\models \phi \}, & \text{if } t\in[t+t_1, t+t_2] \\
    \emptyset & \text{otherwise.}
\end{cases}
\end{align}
along with the corresponding minimal BRT $\mathcal{R}^m(t,T;\mathcal{T}_{\phi}(\cdot))$. Then, $h_{\mathcal{S}_{\square_{[t1,t2]}\phi}}$ can be defined as $\{ x: \not\in \mathcal{R}^m(t,T;\mathcal{T}_{\phi}(\cdot))\}$, i.e., the complement of the minimal BRT.

\section{Proposed framework} \label{proposed_framework}

\subsection{Framework overview}

Throughout this paper, we consider a disturbance-free environment. In other words, we assume $d(\cdot)=0$ in \eqref{dynamics} and \eqref{brt}-\eqref{minimal_brt}\footnote{Although the proposed framework can be applied to a system with disturbances without major modification, it might yield conservative results, i.e., conclude that a feasible mission is infeasible. Addressing such an issue is left for future work.}. 
The objective of the proposed framework is to evaluate whether a given STL specification is feasible, safe, and executable by a target dynamical system, and to synthesize safety controller even under unexpected changes. The proposed framework can largely be divided into two parts: 1) \textit{DeepSTLReach}, a deep reachability-based STL evaluation algorithm, and 2) MILP-based planning and MPC-based control. 

Unlike existing STL-based algorithms or planners, \textit{DeepSTLReach} performs set-based analysis, i.e., it provides a set of states that can satisfy the given STL specification. Thus, one can easily determine whether an arbitrary initial state is feasible for the mission by checking whether it is included in the set, or use this information to select the optimal deployment option according to the preference. 

Once the initial state and a feasible mission plan are chosen, we perform planning and control synthesis using a layered architecture that combines MILP-based global planning at a higher layer of abstraction with MPC-based local planning and control. The main advantage of our proposed architecture is that, even if environmental changes render the global plan infeasible, the local layer can accommodate runtime variabilities. We will highlight the details in the subsequent section.

\subsection{STL evaluation using DeepSTLReach}

Motivated by \cite{chen2018signal}, \textit{DeepSTLReach} aims to deal with the conflict raised by \textit{eventually} or \textit{until} operator by leveraging MRA formulation. Furthermore, we utilize a neural network to reduce the computation burden for the reachability analysis, which can provide stable real-time evaluation performance. 

\subsubsection{Accommodating MRA problem} \label{MRA_section}

Although the feasibility of individual STL subformula can be tested using BRT, this result might introduce a conflict between subformulas. For instance, $\Diamond_{[0,T]} A \wedge \Diamond_{[0,T]} B$ requires  to visit both $A$ and $B$ within time window $[0,T]$. However, \textit{and} operation of the BRTs ($h_{\mathcal{S}_{\Diamond_{[0,T]}A \wedge \Diamond_{[0,T]}B}} = max(h_{\mathcal{S}_{\Diamond_{[0,T]}A}},h_{\mathcal{S}_{\Diamond_{[0,T]}B}})$) might not guarantee the satisfaction of the STL specification, but merely an intersection of BRT for each individual subformula ($\Diamond_{[0,T]}A, \Diamond_{[0,T]}B$). 
While \cite{chen2018signal} resolved this issue using Least-Restrictive Feasible Controller Set (LRFCS), \textit{DeepSTLReach} leverages MRA formulation. Motivated by \cite{chen2025control_arxiv}, \textit{DeepSTLReach} begins by isolating STL subformulas that contain redundancy. 
More specifically, \textit{DeepSTLReach} collects $eventually\ (\Diamond)$ and $until\ (\mathcal{U})$ operators along with the corresponding target and constraint set using equations \eqref{until}-\eqref{always_target}, assuming the target and constraint sets can be rewritten to be disjoint over time\footnote{More specifically, \textit{DeepSTLReach} assumes the given STL specification can be rewritten as separate target-constraint sets that do not impose constraints on one another. Extending the algorithm to general STL specifications will be addressed in future work.}. 

\begin{figure}[t]
    \centering
    \includegraphics[width=0.65\columnwidth]{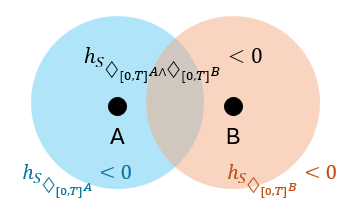}
    \caption{Example of conflict.}
    \label{stl_mra_conflict}
\end{figure}

Assume there exists a total of $N_{MRA}$ target-constraint sets from \textit{eventually} and \textit{until} operators, computed using \eqref{until}. \textit{DeepSTLReach} formulates all the possible ordering of $N_{MRA}$  target-constraint sets and solve the respective MRA problem.
For instance, for an arbitrary sequence of target-constraint sets with orders $\mathbb{T}=( \mathcal{T}_1,\cdots,\mathcal{T}_{N_{MRA}})$ and $\mathbb{C}=( \mathcal{C}_1,\cdots,\mathcal{C}_{N_{MRA}})$, let $h_{\mathcal{T}_i}$ and $h_{\mathcal{C}_i}$ be the functions correspond to $\mathcal{T}_i$ and $\mathcal{C}_i$, respectively. 
Then, \textit{DeepSTLReach} computes the BRT to visit the targets following the sequence $\mathbb{T}$ as follows:
\begin{enumerate}
\item Compute $h_{\mathcal{R}^M_{N_{MRA}}} (t,x)$ where $\mathcal{R}^M_{N_{MRA}} = \mathcal{R}^M_{N_{MRA}}(t,T;\mathcal{T}_{N_{MRA}},\mathcal{C}_{N_{MRA}})$ based on \eqref{brt} and \eqref{Bellman}. In other words, compute the BRT of the last target-constraint set. 
\item For $i=1,\cdots,N_{MRA}-1$, compute $\bar{h}_{\mathcal{T}^M_{N_{MRA}-i}} (t,x)= max(h_{\mathcal{T}^M_{N_{MRA}-i}},h_{\mathcal{R}^M_{N_{MRA}-i+1}}) (t,x)$. Then, compute $h_{\mathcal{R}^M_{N_{MRA}-i}} (t,x)$ where $\mathcal{R}^M_{N_{MRA}-i} =\mathcal{R}^M_{N_{MRA}-i}(t,T;\mathcal{\bar{T}}_{N_{MRA}-i},\mathcal{C}_{N_{MRA}-i})$ with $ \mathcal{\bar{T}}_{N_{MRA}-i}=\{x:\bar{h}_{\mathcal{T}^M_{N_{MRA}-i}} (t,x) < 0\}$, i.e., iteratively computes the BRT backward over the given sequence~\cite{chen2025control_arxiv}.
\end{enumerate}
The last BRT computed from the above process ($\mathcal{R}^M_1$) represents the set of states that can satisfy the STL specifications with the sequential visit of ($\mathbb{T}, \mathbb{C}$). Consequently, one can compute the BRT for each respective sequence. 

Let the computed BRT for ($\mathbb{T}_i, \mathbb{C}_i$) from the above process as $ \mathcal{R}^M_{\mathbb{T}_i, \mathbb{C}_i}$, where ($\mathbb{T}_i, \mathbb{C}_i$) is the $i-$th pair of the sequence among $N_{MRA}!$ cases. Then, one can utilize the union of all the possible orderings of ($\mathbb{T}, \mathbb{C}$), $\bigcup_{i=1} \mathcal{R}^M_{\mathbb{T}_i, \mathbb{C}_i}$, as a corresponding BRT for the collected subformulas with the \textit{eventually} or \textit{until} operator. The remaining subformulas without the \textit{eventually} or \textit{until} operator can be handled by \eqref{corresp}-\eqref{always_target}, as shown in \cite{chen2018signal}.

If a subformula has the nested form, i.e., it contains (sets of) \textit{until} or \textit{eventually} operators within other operators, one can iteratively apply the above process. Although this process might require numerous computations of BRT as $N_{MRA}$ grows, \textit{DeepSTLReach} can efficiently handle this problem using deep learning-based reachability analysis, which will be introduced in the following section. 

\subsubsection{Deep reachability analysis}

To address the computational complexity of solving HJ-VI, 
\textit{DeepSTLReach} approximates the Bellman value function using a deep neural network in a supervised manner~\cite{neural_reach}. 
More specifically, we first solve the HJ-VI using~\cite{chen2018signal} and generate a training dataset $\mathcal{D}
:= \left\{ (t_j, x_k, h_{\mathcal{R}^M}(t_j,x_k)) \right\}$ where $(t_j,x_k)$ spans all points of the discretized state-time grid.
Let $\tilde{h}_\theta(t,x)$ denote a neural approximation of $h_{\mathcal{R}^M}(t,x)$, 
parameterized by a fully-connected multilayer perceptron (MLP) that takes $(t,x)$ as input and outputs a scalar approximation. $\theta$ is trained to minimize a regression loss augmented with a conservative penalty:
\begin{equation} \label{loss_function}
\mathcal{L}(\theta) = \mathcal{L}_{\mathrm{MSE}} + \mathcal{L}_{\mathrm{penalty}}, 
\end{equation}
where
\begin{equation*}
\begin{aligned}
\mathcal{L}_{\mathrm{MSE}}
&=
\mathbb{E}_{(t,x)\sim\mathcal{D}}
\Bigl[
    \bigl(
        \tilde{h}_\theta(t,x)
        - h_{\mathcal{R}^M}(t,x)
    \bigr)^2
\Bigr],
\\
\mathcal{L}_{\mathrm{penalty}}
&=
\lambda\,
\mathbb{E}_{(t,x)\sim\mathcal{D}}
\Bigl[
    \mathbf{1}_{\{h_{\mathcal{R}^M}(t,x)>0\}}
    \max\!\bigl(
        0,\,
        \varepsilon-\tilde{h}_\theta(t,x)
    \bigr)
\Bigr].
\end{aligned}
\end{equation*}
with $\mathcal{L}_{\mathrm{MSE}}$ denoting the supervised regression loss over $\mathcal{D}$, $\mathcal{L}_{\mathrm{penalty}}$ guides toward conservative approximation in unsafe regions to prevent false-safe predictions. 
The penalty discourages underestimation of the value function when $h_{\mathcal{R}^M}(t,x) > 0$, thereby promoting a conservative approximation of the BRT boundary. 
Once trained, the maximal BRT can be evaluated in real time via the sublevel condition $\tilde{h}_\theta(t,x) < 0$, enabling online feasibility monitoring without solving the HJ-VI during runtime. The similar process can be applied for $\mathcal{R}^m$.

\subsection{Layered-architecture for planning and control}

Our proposed layered planning architecture can again be decomposed into two layers: \emph{Global planning via MILP}, which ensures that STL specifications are met by the agent, and \emph{MPC-based local planning and control}, which enables runtime safety assurance even when the original STL and corresponding plan become unsafe during the mission execution.

\subsubsection{Global Planning via Mixed-Integer Linear Programming (MILP)}

Once the STL passes \textit{DeepSTLReach}, we can transform the given STL specification to the constraints of a MILP problem~(\cite{belta2019formal, raman2014model}) via a robustness $\rho$. 
Given the dynamics of the target system \eqref{dynamics}, trajectory length $N$, and STL $\phi$; we denote $x(k|t)$ and $u(k|t)$ as the $k^{\text{th}}$ step forward prediction of the state and inputs at time $t$, respectively. Here, $k \in \{1,\dots, N\}$ and we define $x(0|t) = x(t)$, $\mathbf{u}^{N,t} = [u(0,t)^\top,  \dots, u(N|t)^\top]^\top$, and $\mathbf{x}(x(t), \mathbf{u}^{N,t}) = [x(0,t)^\top, \dots, x(N|t)^\top]^\top$. The optimization problem for motion planning with STL specification as constraints, i.e., for any arbitrary initial states $x(t)$ within the BRT, can be described as:
\begin{align} 
\min_{\mathbf{u}^{N,t}} \quad & J (\mathbf{x}(x(t), \mathbf{u}^{N,t}), \mathbf{u}^{N,t}) \label{mpc1}\\
\text{s.t.} \quad
& \dot{x} = \tilde{f}(x,t, u)\\
& \rho(x(t), \cdots, x(t+N), t, \phi) \geq 0  \label{mpc3}
\end{align}
where $J(\cdot)$ is the cost function that incorporates mission planning and control objectives, and $\tilde{f}$ is linearized dynamics. Let the optimal state trajectories be denoted as $\bar{\mathbf{x}}$. These trajectories will be fed to the MPC-based tracking controller described later.

Equations \eqref{mpc1}-\eqref{mpc3} can be rewritten as a mixed integer formulation by appropriately encoding the predicates of the STL as mixed integer constraints~(\cite{raman2014model}). For each predicate of the form $\pi = (a_\pi^\top x(t) \leq b_\pi)$, we define a variable $z(t)^{\pi}\in \{0,1\}$ such that $1$ (respectively, 0) stands for true (respectively, false).  Dropping the time index and parameters for each of the variables for brevity, the relation between $z^{\pi}$, robustness $\rho$, and state $x$ can be described as $a_{\pi}^\top x - M(1-z^{\pi} + \rho) \leq b_\pi$ and $a_{\pi}^\top x + M z^{\pi} + \rho \geq b_\pi$.
The conversion of a compound STL formula to MILP constraints is performed recursively, which is omitted for brevity (see~(\cite{belta2019formal, raman2014model}) for more details).

\subsubsection{MPC-based Local Planning and Control}
We adopt the MPC framework~\cite{rawlings2017model} to perform local planning and control. The objective is to design a tracking controller that follows the reference trajectory $\bar{\mathbf{x}}$ generated by the MILP planner and also satisfies the STL specification. However, to ensure safety in the presence of runtime safety violations, we employ an MPC with state constraints that enforce system safety, since replanning with MILP can be computationally expensive.
The continuous-time system dynamics are discretized using a fixed sampling period, and the resulting discrete-time model is employed within the MPC framework to compute control inputs at each sampling instant. 

Let $\mathbf{x}_{l|k}$ and $\mathbf{u}_{l|k}$ denote the predicted state and control input at prediction step $l$ computed at time $k$, with $l \in \{0,\dots,N\}$. The prediction horizon length is $N$, and $\mathbf{x}_{0|k} = \mathbf{x}_k$. The stacked control sequence is defined as
$\mathbf{u} = \begin{bmatrix}
u_{0|k}^\top &
u_{1|k}^\top &
\cdots &
u_{N-1|k}^\top
\end{bmatrix}^\top$.
The MPC optimization problem is formulated as: 
\begin{subequations}
\label{eq:mpc_single}
\begin{align}
J^*(\mathbf{x}_k)
=
\min_{\mathbf{u}}
&\left\| \mathbf{x}_{N|k} - \bar{\mathbf{x}}_{N|k} \right\|_{P}^{2} + \nonumber \\
& \sum_{l=0}^{N-1}\left\| \mathbf{x}_{l|k} - \bar{\mathbf{x}}_{l|k} \right\|_{Q}^{2}
+ \left\| \mathbf{u}_{l|k} - \mathbf{u}_{l-1|k} \right\|_{R}^{2}
\label{eq:mpc_cost_single} \\
\text{s.t.} \quad
& \mathbf{x}_{0|k} = \mathbf{x}_k, \label{eq:mpc_init_single} \\
& \mathbf{x}_{l+1|k} = \bar{f}(\mathbf{x}_{l|k},\mathbf{u}_{l|k}) \forall l = 0,\dots,N-1, \label{eq:mpc_dyn_single} \\
& (\mathbf{x}_{l|k},\mathbf{u}_{l|k}) \in \mathcal{Z}, \label{eq:mpc_const_single} \\
& \mathbf{x}_{N|k} \in \Gamma, \label{eq:mpc_terminal_single} \\
& h^j(\mathbf{x}_{l|k},\mathbf{u}_{l|k}) \geq 0, \quad \forall j = 1,\dots,m. \label{eq:mpc_cbf_single}
\end{align}
\end{subequations}
where $P$, $Q$, and $R$ are positive-definite weighting matrices of appropriate dimensions. $\bar{f}(\cdot)$ denotes the discrete-time linearized dynamics, $\mathcal{Z}$ represents admissible state and input constraints, and $\Gamma$ is a terminal invariant set; this can be computed using a control-invariant set of a linearized controller at the terminal condition. Finally, $h^j(\cdot)$ denotes the safety constraints (e.g., collision avoidance constraints).
This formulation ensures recursive feasibility, reference-tracking performance, and provable safety by integrating state constraints. It is worth noting that we assume that any change in the constraints is detected early enough that a feasible trajectory exists over the entire prediction horizon at the time of detection.
For example, a warehouse robot's perception layer must detect an approaching obstacle while a collision-free predicted trajectory exists at the current time step.
\begin{rem}
The result of \textit{DeepSTLReach} can be easily integrated as a safety constraint $h^j(\cdot)$ in \eqref{eq:mpc_cbf_single} as the basis of the neural CBFs (\cite{so2024train}) or control barrier value functions (\cite{choi2021robust}). Such an extension will be investigated as future work. 
\end{rem}

\section{Numerical Simulation} \label{sec:experiment}

\subsection{Simulation setup}
In this section, we validate the effectiveness of our proposed framework using illustrative numerical simulations. Throughout the simulations, we set the state and input of a target dynamical system as 
\begin{equation}
    x=[p_x,p_y,\dot{p}_x,\dot{p}_y]^T, \ u=[\ddot{p}_x, \ddot{p}_y]^T
\end{equation}
where $p_x$ and $p_x$ are the x and y-axis position, respectively. For the sake of simplicity, we assume the dynamics of the system as a double integrator. 
We also assume that environmental information (e.g., obstacle locations, waypoints, goal points) is known and available to all components of the framework, including BRT computation. For instance, obstacles can always be included in the constraint set when a BRT is generated.

For \textit{DeepSTLReach}, we train a network of four hidden layers with 128 neurons per layer and ReLU activation functions. The expectations in \eqref{loss_function} are approximated using mini-batch stochastic gradient descent with the Adam optimizer. We set $\epsilon = 0.1$ and $\lambda = 0.05$ for the training. 

All the computation times for the simulation scenarios are obtained using a computer equipped with an AMD Ryzen 7 8845HS processor and an NVIDIA RTX 4050 GPU. We implement each of our mission planning scenarios using the \texttt{stlpy} library~\cite{kurtz2022mixed} and the Gurobi solver~\cite{gurobi}, along with \texttt{stlcg} library~\cite{leung2023backpropagation}. The baseline BRT is computed by utilizing \texttt{hj-reachability}~\cite{github}. 

\subsection{Scenario 1: Unexpected moving obstacle} \label{scen1}
\begin{figure}[t]
    \centering

    \begin{minipage}[b]{0.49\linewidth}
        \centering
        \includegraphics[width=0.85\linewidth]
        {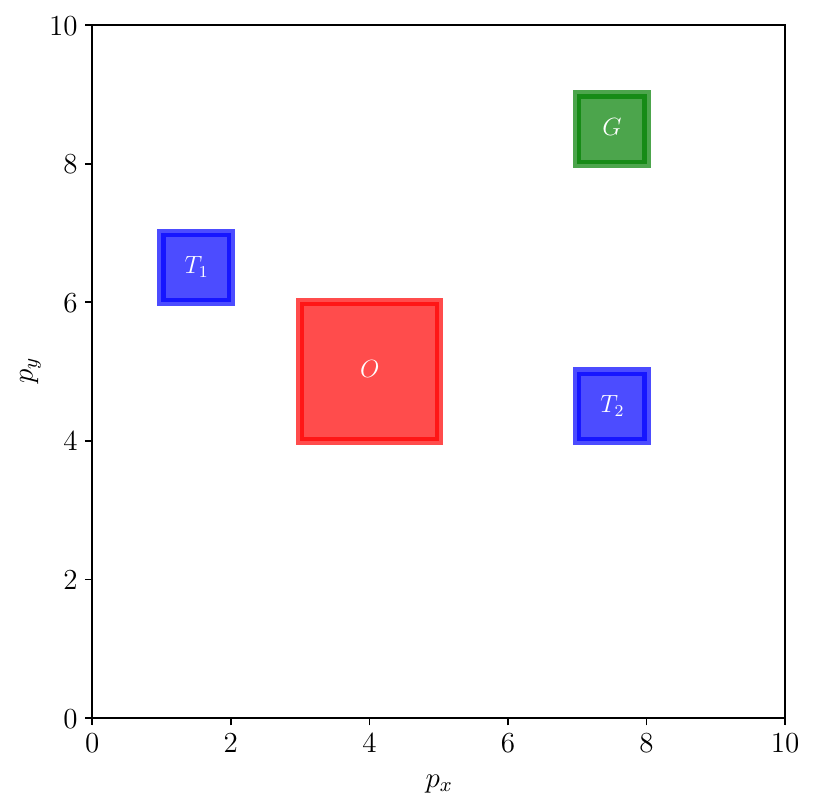}
    \end{minipage}
    \begin{minipage}[b]{0.49\linewidth}
        \centering
        \includegraphics[width=\linewidth]
        {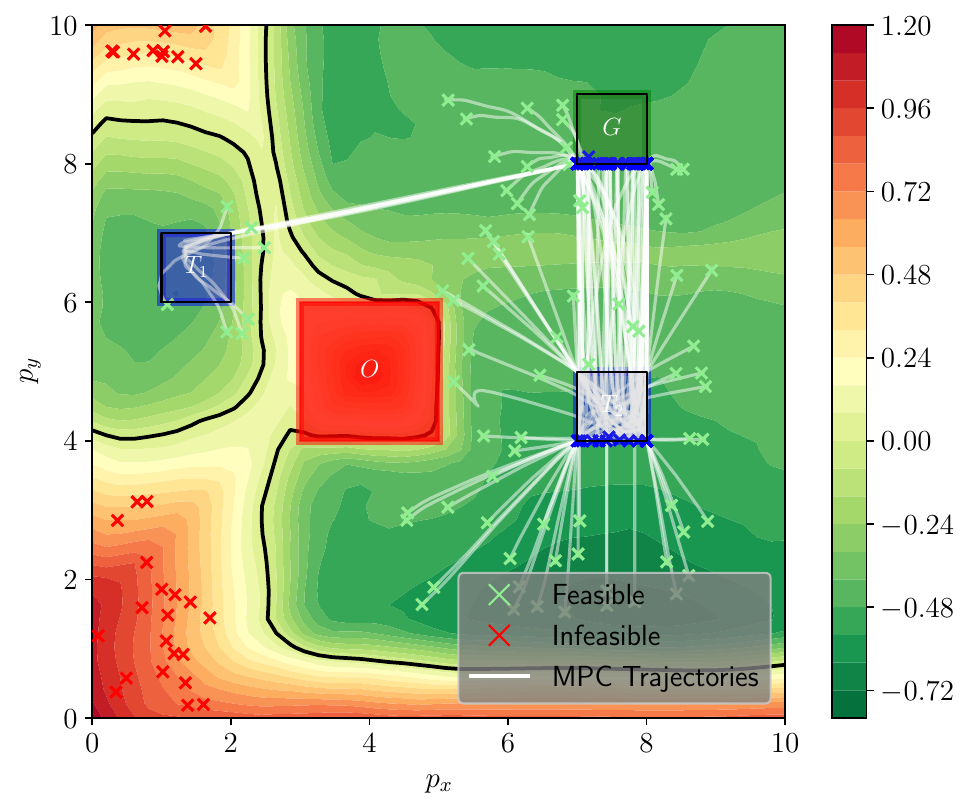}
    \end{minipage}

    \vspace{0.5em}

    \begin{minipage}[b]{0.49\linewidth}
        \centering
        \includegraphics[width=0.85\linewidth]
        {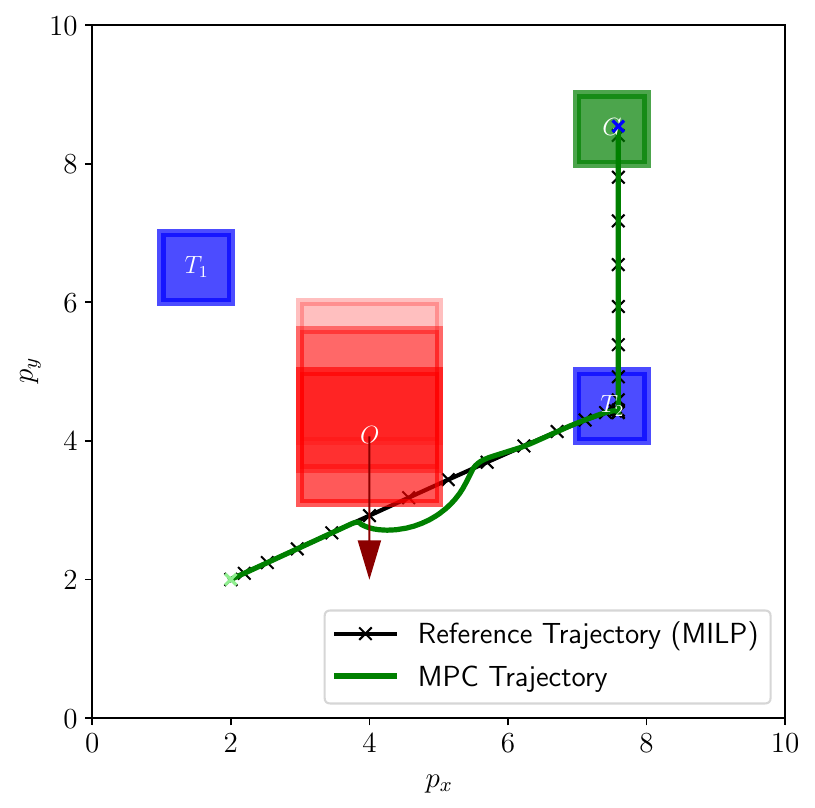}
    \end{minipage}
        \begin{minipage}[b]{0.49\linewidth}
        \centering
        \includegraphics[width=0.85\linewidth]
        {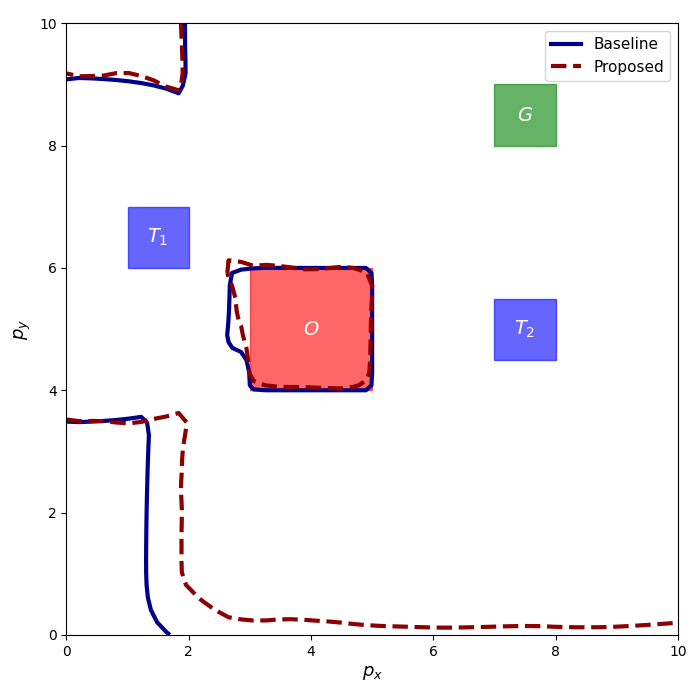}
    \end{minipage}

    \caption{Simulation environment and results of Scenario 1.
    (\textit{Top left}) Scenario 1 map.
    (\textit{Top right}) Simulation result with $T=8\,\mathrm{s}$.
    (\textit{Bottom left}) Runtime assurance of MPC under unexpected
    environment changes. 
    (\textit{Bottom Right}) Comparison of BRT between baseline and proposed framework with $T=9\,\mathrm{s}$.}
    \label{scen1_result}
\end{figure}

In this section, we present the simulation results based on the scenario borrowed from \cite{kurtz2022mixed} (Simulation Experiments (a)). For this scenario, we set $p_x,p_y\in[0,15]$, $|\dot{p}_x|\leq1$, $|\dot{p}_y|\leq1$, $|\ddot{p}_x|\leq 0.5$, and $|\ddot{p}_y|\leq 0.5$.
The STL specification is given as:
\begin{equation} \label{scen1_stl}
    \Diamond_{[0,T-5]}(\square_{[0,5]}T_1 \vee \square_{[0,5]}T_2) \wedge \square_{[0,T]} \neg O \wedge \Diamond_{[0,T]}G, 
\end{equation}
where $T$ is the mission time.  The top left figure of Fig.~\ref{scen1_result} shows the map of Scenario 1. There exist two waypoints ($T_1$ and $T_2$) represented as the blue boxes, one goal point ($G$, green), and one obstacle ($O$, red). 

The top right figure of Fig.~\ref{scen1_result} shows the simulation results. In the figure, the black line denotes the zero-level set boundary of the computed BRT; the green $x$ denotes the feasible initial states; the white line denotes the trajectory generated by the MPC-based controller \eqref{eq:mpc_cost_single}; and the blue $x$ denotes the final state of the trajectories. The red $x$ denotes infeasible initial states, i.e., the MILP planner \eqref{mpc1} cannot find a valid trajectory that satisfies the given STL \eqref{scen1_stl}. 
As shown the figure, the computed BRT from \textit{DeepSTLReach} successfully identifies the infeasible region to perform the mission \eqref{scen1_stl}. 
Monte Carlo simulations with 100 randomly sampled initial states show that all states outside the BRT fail to find a feasible trajectory, whereas all states within the BRT succeed in finding a feasible trajectory using the proposed MPC controller.

Meanwhile, the bottom left of Fig.~\ref{scen1_result} corresponds to the case in which the obstacle ($O$) moves downward, i.e., an unexpected environmental change occurs (such that MPC can detect this change while a feasible path exists). As a result, the original plan generated by the MILP planner becomes infeasible, since its trajectory collides with the obstacle. Nevertheless, the proposed MPC controller accounts for environmental changes while satisfying the given STL specification, thereby enabling robust mission execution.

The bottom right figure shows the comparison of computed BRT using the baseline and proposed framework. In the figure, the blue line represents the result from the baseline method, and the red dashed line is the result from the proposed algorithm. Despite some differences, one can observe that the computed BRTs show similar boundaries. Furthermore, the \textit{DeepSTLReach} generates more conservative (smaller) BRT, showing the effect of the designed penalty function \eqref{loss_function}.

\begin{table}[t]
    \centering
    \caption{BRT computation time (\(\mathrm{s}\))}
    \label{scen_computation}

    \setlength{\tabcolsep}{4pt}

    \begin{tabular}{lcc}
        \toprule
          & Baseline & \textbf{Proposed} \\
        \midrule
        Inference time          & 39.488         & \textbf{0.038}   \\
        \bottomrule
    \end{tabular}
\end{table}



Table~\ref{scen_computation} shows the BRT inference time for Scenario 1 with $T=10[s]$. In the case of \textit{DeepSTLReach}, it includes the time to compute all the necessary BRTs to obtain $\bigcup_{i=1} \mathcal{R}^M_{\mathbb{T}_i, \mathbb{C}_i}$, not just a single BRT. 
For the first column (Baseline), we use \textit{hj\_reachability} to compute BRT, whereas the second column (Proposed) uses the trained neural network with loss function \eqref{loss_function}. All the times are computed as the average of three independent computations. 
As shown in the table, the trained neural network achieves significantly faster performance, demonstrating the efficiency of the proposed framework.

\subsection{Scenario 2: Door and key} \label{scen3_appendix}

\begin{figure}[h] 
    \centering
    \begin{minipage}{0.35\textwidth} 
        \centering
        \includegraphics[width=0.95\linewidth]{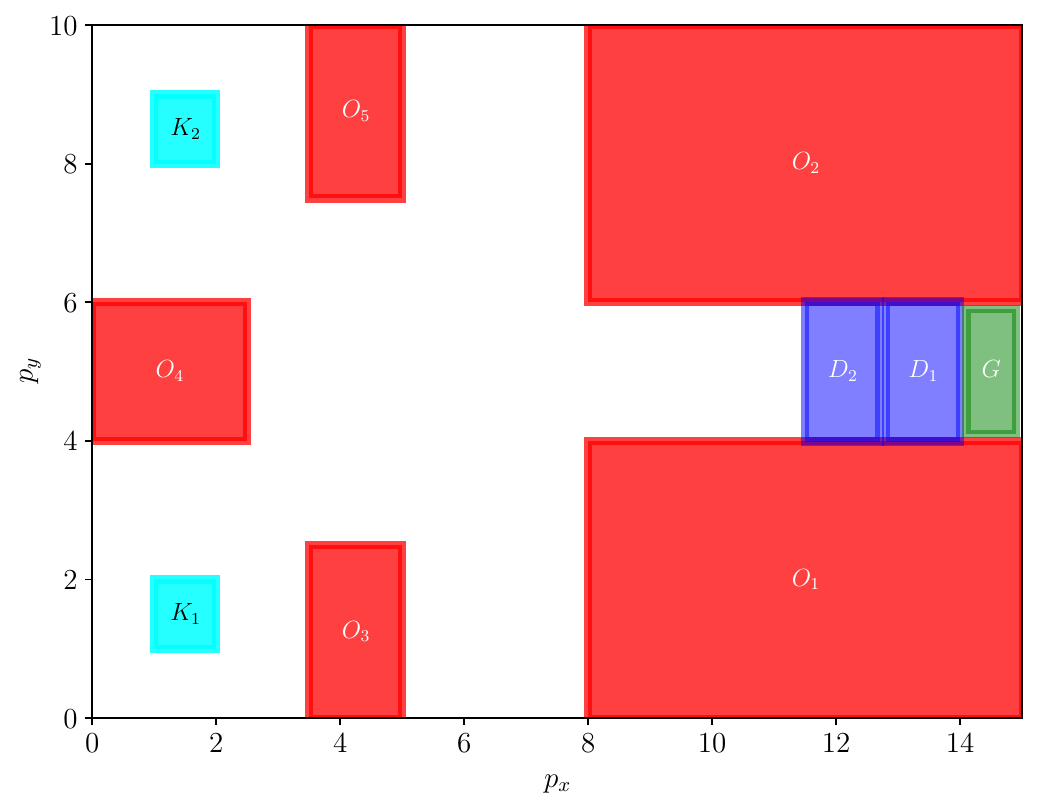}
    \end{minipage}
    \begin{minipage}{0.4\textwidth}
        \centering
        \includegraphics[width=\linewidth]{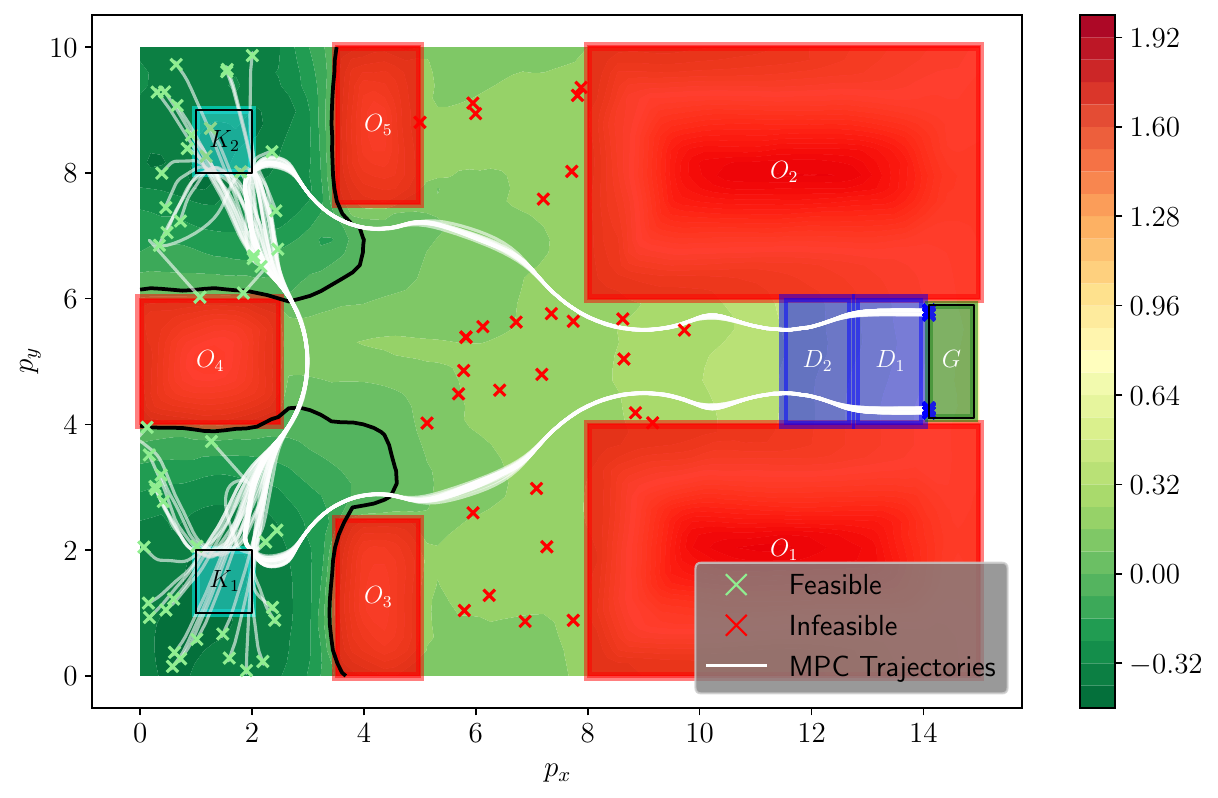}
    \end{minipage}
    \caption{Simulation environment and results of Scenario 2. (\textit{Top}) Scenario 2 map. (\textit{Bottom}) Simulation result with $T=20s$.}
    \label{scen3_result}
\end{figure}

Similar to the previous simulation, Scenario 2 is set based on \cite{kurtz2022mixed} (Simulation Experiment (d)). The STL specification given for Scenario 2 is defined as
\begin{equation}
   \bigwedge_{i=1}^2 (\neg D_i \mathcal{U}_{[0,T]}K_i) \wedge \Diamond_{[0,T]}G \wedge \square_{[0,T]}(\bigwedge_{i=1}^5 \neg O_i)
\end{equation}
where $K_i$ represents the key for door $D_i$, i.e., $D_i$ opens only when the target system visits the corresponding $K_i$. The detailed map of Scenario 2 can be found in Fig.~\ref{scen3_result}. For this scenario, we set $p_x\in[0,15]$, $p_y\in[0,10]$ $|\dot{p}_x|\leq2$, $|\dot{p}_y|\leq2$, $|\ddot{p}_x|\leq 0.5$, and $|\ddot{p}_y|\leq 0.5$. All other parameters are identical to those of the previous experiment.
Similar to the previous simulation, the proposed framework successfully estimates the BRT with respect to the given STL, generating a valid trajectory using the proposed MPC controller to accomplish the mission.





\section{Conclusion} \label{sec:conclusion}

In this paper, we propose an STL-based framework for evaluating feasibility and executing STL missions. \textit{DeepSTLReach} evaluates the specification by computing the BRT via deep reachability analysis, providing comprehensive feasibility information, including \textit{where to start the mission}, rather than evaluating only a specific initial state. Furthermore, our layered architecture combines global and local planning to ensure system safety even if the initial plan's feasibility changes during execution. Numerical simulations validate the proposed framework's ability to efficiently evaluate STL specifications. Future work will address the scalability of \textit{DeepSTLReach} by employing semi-supervised learning techniques. Although \textit{DeepSTLReach} significantly improves the inference time, it still requires some time to generate datasets, which we plan to address by leveraging semi-supervised learning. We also plan to resolve the expression limitations of \textit{DeepSTLReach}, such as accommodating the existence of disturbances or globally correlated target-constraint sets. 

\bibliographystyle{plain}
\bibliography{Neus_bib}


\end{document}